\documentclass[pdflatex,sn-mathphys]{sn-jnl}

\usepackage[version=4]{mhchem}

\jyear{2023}%

\theoremstyle{thmstyleone}%
\theoremstyle{thmstyletwo}%

\theoremstyle{thmstylethree}%

\usepackage{hyperref}

\usepackage{cleveref}

\renewcommand{\deg}{^\circ}

\newcommand{\au}{\,\mathrm{au}}
\newcommand{\km}{\,\mathrm{km}}
\newcommand{\meter}{\,\mathrm{m}}

\newcommand{\mm}{\,\mathrm{mm}}
\newcommand{\um}{\,\mu \mathrm{m}}
\newcommand{\nm}{\,\mathrm{nm}}

\newcommand{\Myr}{\,\mathrm{Myr}}
\newcommand{\kyr}{\,\mathrm{kyr}}

\newcommand{\days}{\,\mathrm{d}}

\newcommand{\hour}{\,\mathrm{h}}

\newcommand{\second}{\,\mathrm{s}}

\newcommand{\K}{\,\mathrm{K}}
\newcommand{\J}{\,\mathrm{J}}

\newcommand{\kg}{\,\mathrm{kg}}

\newcommand{\mol}{\,\mathrm{mol}}
\newcommand{\tiu}{\,\mathrm{J}\ \mathrm{K}^{-1} \mathrm{m}^{-2}\mathrm{s}^{-1/2}}

\begin{document}

\title[Thermal decomposition activity of Phaethon]{Thermal decomposition as the activity driver of near-Earth asteroid (3200) Phaethon}


\author*[1]{\fnm{Eric} \sur{MacLennan}}\email{eric.maclennan@helsinki.fi}

\author[1,2]{\fnm{Mikael} \sur{Granvik}}\email{mgranvik@iki.fi}

\affil*[1]{\orgdiv{Department of Physics}, \orgname{University of Helsinki}, \orgaddress{\city{Helsinki}, \country{Finland}}}

\affil[2]{\orgdiv{Asteroid Engineering Laboratory}, \orgname{Lule\aa\ University of Technology}, \orgaddress{\city{Kiruna}, \country{Sweden}}}



\abstract{Near-Earth asteroid (3200) Phaethon exhibits activity during its perihelion passage at 0.14 au from the Sun and is the likely parent body of the annual Geminid meteor shower. Its low albedo and featureless B-type reflectance spectrum indicate a primitive composition, but a definitive meteorite analog is currently indeterminate. Here, we analyze a mid-infrared emissivity spectrum of Phaethon and find that it most closely matches the Yamato-group (CY) of carbonaceous chondrites. The CY chondrites experienced aqueous alteration and recent thermal metamorphism in which extreme temperatures caused mineral decomposition, resulting in the production of gas species. Temperatures within Phaethon during its close approach to the Sun are conducive to the thermal decomposition of carbonates, Fe-sulfides, and phyllosilicates that release CO$_2$, S$_2$, and H$_2$O gas, respectively. Spectral detection of these minerals strongly implies that gas release from mineral decomposition is capable of triggering dust ejection. The planned flyby of Phaethon by the DESTINY$^+$ spacecraft in 2028 will allow us to verify this hypothesis.}

\maketitle

\section*{ }\label{sec1}

The near-Earth asteroid (NEA) (3200) Phaethon is one of the few objects classified as an active asteroid \citep{Jewitt12}, due to an active tail that has consistently been observed during the last three perihelion passages \citep{Jewitt&Li2010,Jewitt_etal2013,Li&Jewitt2013,Hui&Li2016}. Phaethon's activity and orbital proximity to an observed dust stream strongly suggest that it is the parent body of the annual Geminid meteor shower \citep{Whipple83,Williams&Wu1993,Battams_etal22}. Several plausible activity mechanisms have been proposed: volatilization/sublimation of rock, thermal stress fracturing, meteoroid impacts, and thermal radiation pressure \citep{Jewitt&Li2010,Szalay_etal19,Wiegert_etal20,Masiero_etal21,Bach+Ishiguro21,MacLennan_etal21,Lisse+Steckloff22}. Although the plausibility of many of these processes is evident assuming appropriate circumstances, the ambiguity in Phaethon's composition prevents us from either accepting or rejecting any of the above proposed mechanisms.

\begin{figure}[h!]
	\centering
	\includegraphics[width=0.95\linewidth]{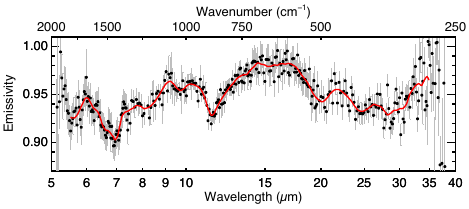}
	\caption{The modeled mid-infrared emissivity spectrum of Phaethon, calculated from fluxes observed by Spitzer Space Telescope's Infrared Spectrograph (IRS) on January 14, 2005. Vertical grey lines indicate the $1\sigma$ uncertainty for each emissivity value. The red curve is a best-fit \texttt{loess} function (\autoref{sec3}).}\label{fig1}
\end{figure}

Phaethon's low geometric albedo \citep[$p_V\approx11\%$;][]{Geem_etal22} and B-type reflectance spectrum are consistent with carbonaceous chondrite (CC) meteorites \citep{Licandro_etal2007,Clark_etal10,deLeon_etal12,Lazzarin_etal19,Kareta_etal21}. While the Karoonda group (CK) of CCs has been proposed as a spectral analog \citep{Clark_etal10,deLeon_etal12}, others have suggested heated CI (Ivuna group), CM (Mighei group) chondrites as well as a few unusual CCs \citep{Licandro_etal2007,Lazzarin_etal19,Kareta_etal21}. A handful of these unusual CCs have recently been recognized as belonging to a newly proposed carbonaceous chondrite group: the CY (Yamato group) chondrites \citep{Ikeda92,King_etal19}. Identifying an appropriate meteorite analog group for Phaethon is currently limited due to a lack of diagnostic absorption features in its visible and near-infrared reflectance spectrum. However, because mid-infrared (mid-IR) emission spectroscopy is sensitive to the bending and stretching frequency modes of minerals that are present on planetary bodies \citep{Salisbury_etal92,Hamilton10,Lane_etal11}, it is particularly useful in compositional studies of planetary bodies with featureless reflectance spectra. By analyzing previously-published mid-IR observations of Phaethon \citep{Hanus_etal2016} from the Spitzer Space Telescope, we calculate Phaethon's mid-IR emissivity spectrum (\autoref{fig1}), search for a spectral meteorite analog, and model the presence of various mineral species.

\section*{Results}

\begin{figure}[h!]
	\centering
	\includegraphics[width=0.95\linewidth]{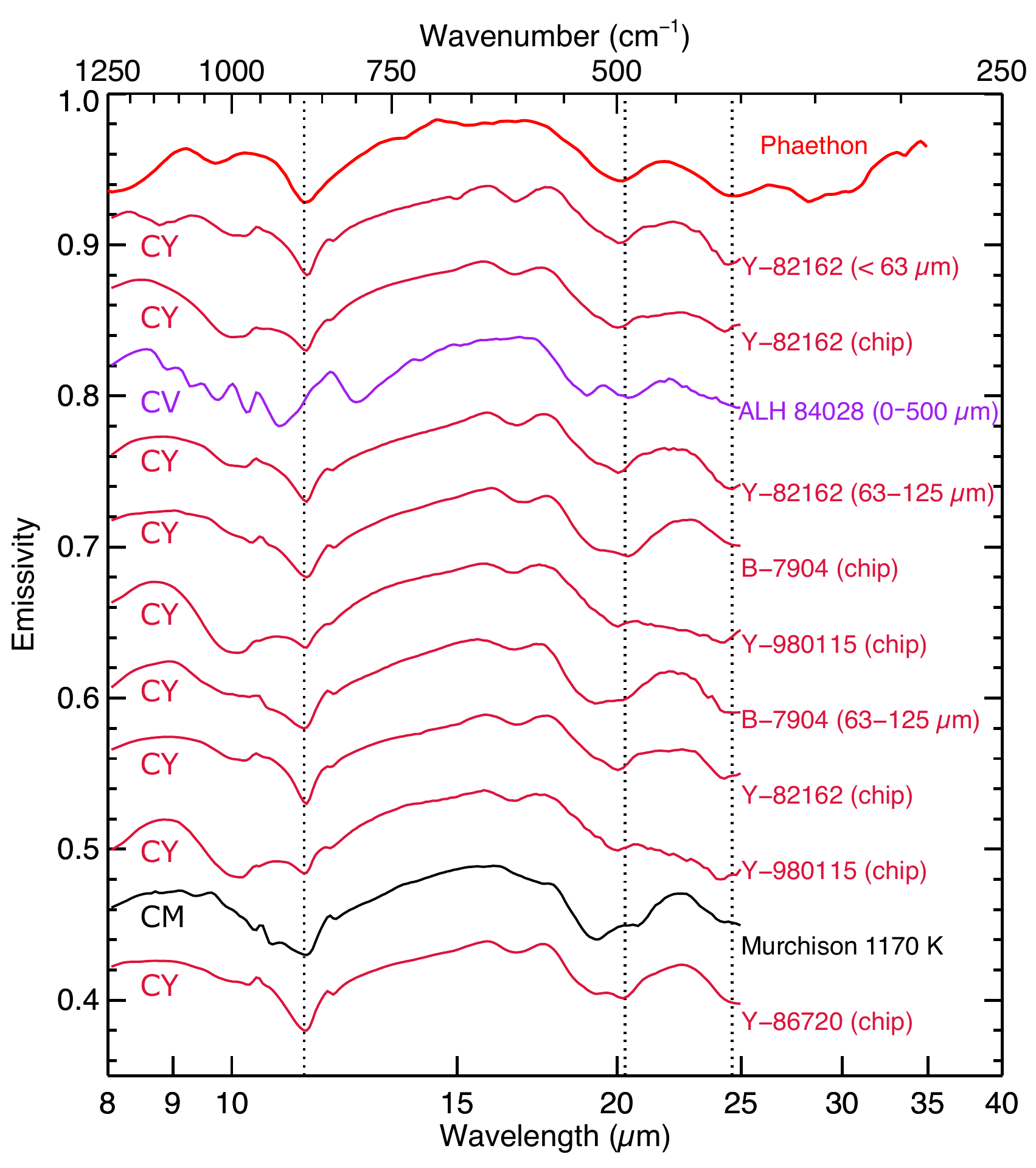}
	\caption{Spectral comparison between Phaethon (top, red) and various meteorite samples. The meteorite name is listed next to its spectrum, with the particulate size (in microns) or chip (centimeter-scale sample) given in parentheses. Vertical dotted lines indicate prominent emissivity features common to Phaethon and the CYs. }\label{fig2}
\end{figure}

\paragraph{Comparison to Meteorite Spectra}
We search for meteorite analogs of Phaethon by using a principal-component analysis (PCA) of meteorite laboratory spectra (\autoref{sec3}). The Vigarano group (CV), CK, Renazzo-type (CR), and Ornans group (CO) can be ruled out by visual inspection of the principal components (PCs), yet many CI, CM, and CY meteorites bear closer resemblance to Phaethon's spectrum (Extended Data \autoref{figex1}). We search for closely-matching samples by calculating differences in PCs between these meteorites and Phaethon. Based on a dissimilar $11.4\um$ feature, we rule out ALH 84028 (CV) as a possible match. Additionally, a Murchison (CM) sample that was experimentally heated to $\sim1170\K$ provide a good spectral match to Phaethon, although dissimilar emissivity features near 21$\um$ effectively eliminate it as a potential spectral analog. Nine of the eleven meteorite spectra most similar to Phaethon's spectrum are CY meteorites and include all of the CY spectra considered (\autoref{fig2}). Out of these spectra, the top three are derived from the same meteorite sample: Y-82162.

The CY chondrites are a scarce (less than 10) group of CCs that experienced aqueous alteration ($<470\K$) for $< 10\Myr$, which resulted in the formation of phyllosilicates, and recent late-stage heating ($>770\K$) \citep{Ikeda92,King_etal19}. Short-duration (1$\hour - 1000\days$) heating resulted in the alteration and formation of various mineral species; notably, the dehydration of phyllosilicates and subsequent transformation into secondary olivine \citep{Nakamura05,Nakato_etal08,King_etal19,Lindgren_etal20}. Such mineralogical changes correspond to various peak metamorphic temperatures, which can be classified into various heating stages \citep{Nakamura05,King_etal21t}: Stage I ($<570\K$), II ($570\K-770\K$), III ($770\K-1020\K$), and IV ($>1020\K$). Many members of the CY group have been classified as stage III or IV, in which most phyllosilicates recrystallize into secondary olivine ($670\K-770\K$) \citep{Nakamura05,Nakato_etal08,Matsuoka_etal22,King_etal19,King_etal21t}. The best-match spectral analog to Phaethon in the mid-IR, Y-82162, has been assessed as stage III \citep{King_etal15a,King_etal15,King_etal19} which implies heating in the range $770\K-1020\K$. Finally, the Murchison sample that was heated to $1170\K$ for a few days has similar spectral properties to Phaethon, which is heated to comparatively lower temperatures ($770\K-1020\K$) for a few weeks near its perihelion \citep{MacLennan_etal21}.

Although the CY meteorites exhibit similar characteristics to other naturally-heated CCs, a separate classification is warranted owing to a few distinct properties \citep{Nakamura05,Matsuoka_etal22,King_etal19,Suttle_etal21,King_etal21t}. In particular, the CY chondrites are distinguished by their heavier oxygen isotopes, relatively higher Fe-sulfide abundance ($20\%-30\%$), low magnetite (Fe$^{2+}$Fe$^{3+}_2$O$_4$) abundance ($\sim1\%$), and the presence of metal \citep{King_etal19,Suttle_etal21}. Dehydrated, dehydroxylated phyllosilicates and/or secondary olivine together comprise $60-70\,$vol.\% of the CYs \citep{King_etal19}. Carbonates are in trace abundance: only a few small Mg-Fe carbonate grains can be found in a few CYs, yet have been estimated to comprise $5\,$vol.\% of the recently proposed CY chondrite Dhofar 1988 \citep{Suttle_etal21}. Fe-sulfides in the CYs are sulfur-depleted, and close to troilite (FeS) in composition, indicating reducing conditions during thermal alteration \citep{Schrader_etal21,Suttle_etal21}. Next, we seek to detect the presence of minerals common to the CYs in Phaethon's spectrum using a spectral mixing model.

\paragraph{Spectral Linear Mixing Model}
The emissivity spectrum of Phaethon is modeled as a linear combination of minerals that are common to carbonaceous chondrites and their associated alteration products (\autoref{sec3}). We determined the best-fit solution to be a mixture of Mg-rich olivine (Fo$_{75}$), troilite (FeS), calcite (CaCO$_3$), magnesite (MgCO$_3$), and the hydroxides portlandite, Ca(OH)$_2$, and brucite, Mg(OH)$_2$. The presence and abundances of these minerals are summarized in Table 1 strongly supports the connection between Phaethon and the CY chondrites. We find that because particle sizes influence the spectral contrast of the samples \citep{Salisbury_etal92,Hamilton10,Lane_etal11}, the uncertainties in model abundances may be larger than those reported in Table 1. In particular, the abundance of troilite may be overestimated due to its smaller spectral contrast relative to olivine (see \autoref{sec3}).

\begin{figure}[h!]
	\centering
	\includegraphics[width=0.95\linewidth]{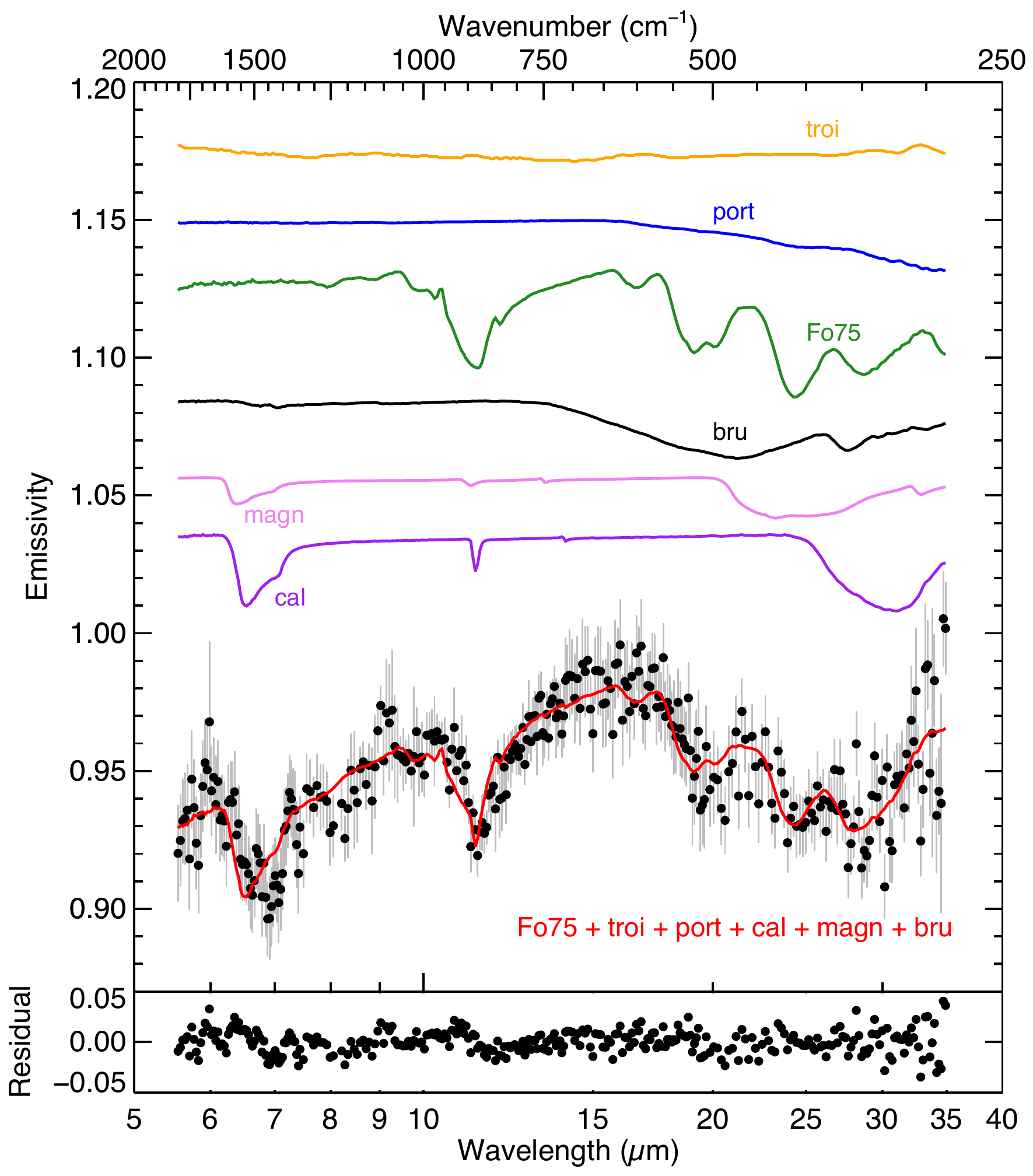}
	\caption{Best-fit spectral mixing model spectrum (red) lain overtop Phaethon's emissivity spectrum (as in Figure 1) and individual component spectra: troilite (orange, troi, FeS), Fo$_{75}$ olivine (green, ol, Mg$_{1.5}$Fe$_{0.5}$SiO$_4$), brucite (black, bru, Mg(OH)$_2$), portlandite (blue, port, Ca(OH)$_2$), calcite (purple, cal, CaCO$_3$), and magnesite (pink, magn, MgCO$_3$).}\label{fig3}
\end{figure}

A sharp 11.4$\um$ feature is characteristic among spectra of naturally and artificially heated CCs and is indicative of olivine that is recrystallized from phyllosilicate as a result of this heating \citep{Lindgren_etal20,Bates_etal21}. In Phaethon's case, the shape of this feature is a consequence of overlapping olivine and carbonate features (green, purple, and pink curves in \autoref{fig3}). The broad emissivity feature in the volume scattering region ($5\um-8\um$) is associated with carbonate cation-oxygen bonds within carbonates (Ca-O and Mg-O for calcite and magnesite, respectively) and their corresponding hydroxides (portlandite and brucite). We found that replacing calcite and magnesite with other carbonates such as siderite (FeCO$_3$) yielded a comparable, but slightly worse, fit (Extended Data \autoref{extab1}). At longer wavelengths, vibrational features consistent with Mg-rich olivine are clearly present near $20\um$, $25\um$, and $28\um$ \citep{Hamilton10}. We do not find evidence of phyllosilicate features in the mid-IR spectrum; a result which is consistent with both the lack of hydration features in reflectance spectra at visible and near-infrared wavelengths \citep{Takir_etal20} and with our PCA model results that indicate a dissimilarity between Phaethon and the phyllosilicate-rich CI and CM chondrites \citep{King_etal15}.

\paragraph{Thermal Metamorphism of CY meteorites}
Late stage heating in CY chondrites is evidenced by partly decomposed phyllosilicates, carbonates, and sulfides \citep{Tonui_etal03,Tonui_etal14,King_etal19}. Specifically, phyllosilicates are transformed into secondary olivine at $870\K-970\K$ after they have undergone dehydration and dehydroxylation at lower temperatures ($300\K-500\K$). Laboratory heating of CIs and CMs show that water vapor is produced in this process \citep{Nakamura05,Nakato_etal08,Tonui_etal14,King_etal15,King_etal21h}. The best-fit spectral mixing model indicates a significant abundance of Mg-rich olivine on Phaethon ($\sim36\%$), which is consistent with the decomposition of phyllosilicates due to extreme heating.

While we do not find evidence of periclase, or lime (CaO) in Phaethon's spectrum, the portlandite and brucite could have formed from the rehydration of these decomposition products \citep{Haberle+Garvie17} (see also the Discussion). Although carbonates are uncommon among CY meteorites, absorption features near 7$\um$ is evidence that they are present at Phaethon's surface, and are  likely a result of early aqueous alteration in the parent body \citep{King_etal19}. Periclase (e.g., MgO and FeO) and other metal-oxide clasts found in a few CYs \citep{Tonui_etal03,Tonui_etal14} have been proposed as decomposition products of magnesite (MgCO$_3$), siderite (FeCO$_3$), and other carbonates \citep{Tonui_etal03,King_etal15a}. Therefore, we suppose that carbonates detected in Phaethon's spectrum represent precursors of these unique clasts found in the CYs \citep{Tonui_etal03,King_etal19}.

Naturally-heated CMs, and experimentally heated CI and CM samples, show evidence of sulfur depletion with increased heating \citep{Tonui_etal14,Schrader_etal21} that leads to higher amounts of troilite as S is lost from Fe-poor sulfide (i.e., FeS$_2$). Sulfur gas was detected during experimental heating of ALH 83100 (CM) in the range $670\K-1070\K$, corresponding to the appearance of unknown Ca-rich sulfide \citep{Lindgren_etal20}. Petrographic analyses of CYs suggest that sulfur gas was released from Fe-sulfide decomposition \citep{Haberle+Garvie17} (see \autoref{eq:sulf}).

\begin{figure}[h!]
	\centering
	\includegraphics[width=0.95\linewidth]{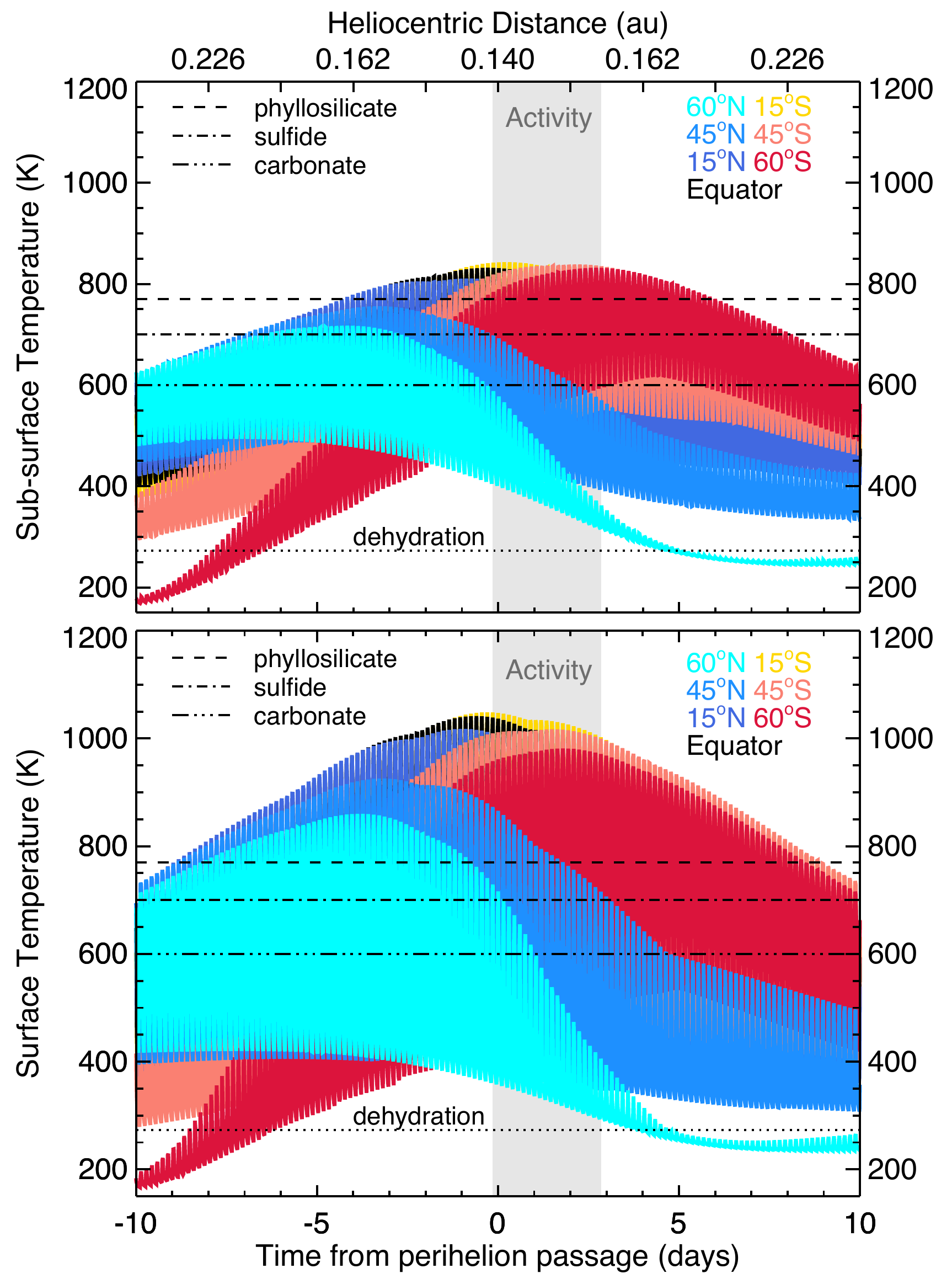}
	\caption{Temperatures estimated at one diurnal thermal skin depth ($l_s$) beneath the surface (top) and at the surface (bottom) Phaethon during perihelion passage. Note that Phaethon's rotation every $\sim3.6 \hour$ causes temperature variation at each latitude, which are colored according to the legend. The lower temperature range at which gas species are produced via thermal decomposition are indicated as horizontal lines.}\label{fig4}
\end{figure}

\paragraph{Thermal Decomposition on Phaethon}
The temperatures at which the thermal decomposition of phyllosilicates, carbonates, and sulfides occur are consistent with Phaethon's perihelion temperatures \citep{MacLennan_etal21}. Here, we model the temperatures of Phaethon and compare them to the temperature ranges over which gas release has been observed in experimental heating of carbonaceous chondrites \citep[i.e.,][]{King_etal15}. Surface and sub-surface temperatures are consistent with phyllosilicate, carbonate, and sulfide decomposition for $\pm 10\days$ around perihelion \autoref{fig4}. The sub-surface temperatures are shown for one thermal skin depth ($l_s$) beneath the surface, which is approximately $l_s = 34.5^{+6.3}_{-4.1} \mm$ at a heliocentric distance of 1.125$\au$ \citep{MacLennan_etal22} and increases with temperature \citep{Rozitis_etal2018}. We use these temperatures and laboratory measurements of reaction-rate constants to estimate the gas pressures and production rates expected from each reaction mechanism on Phaethon (see \autoref{sec3}).

\begin{figure}[h!]
	\centering
	\includegraphics[width=0.95\linewidth]{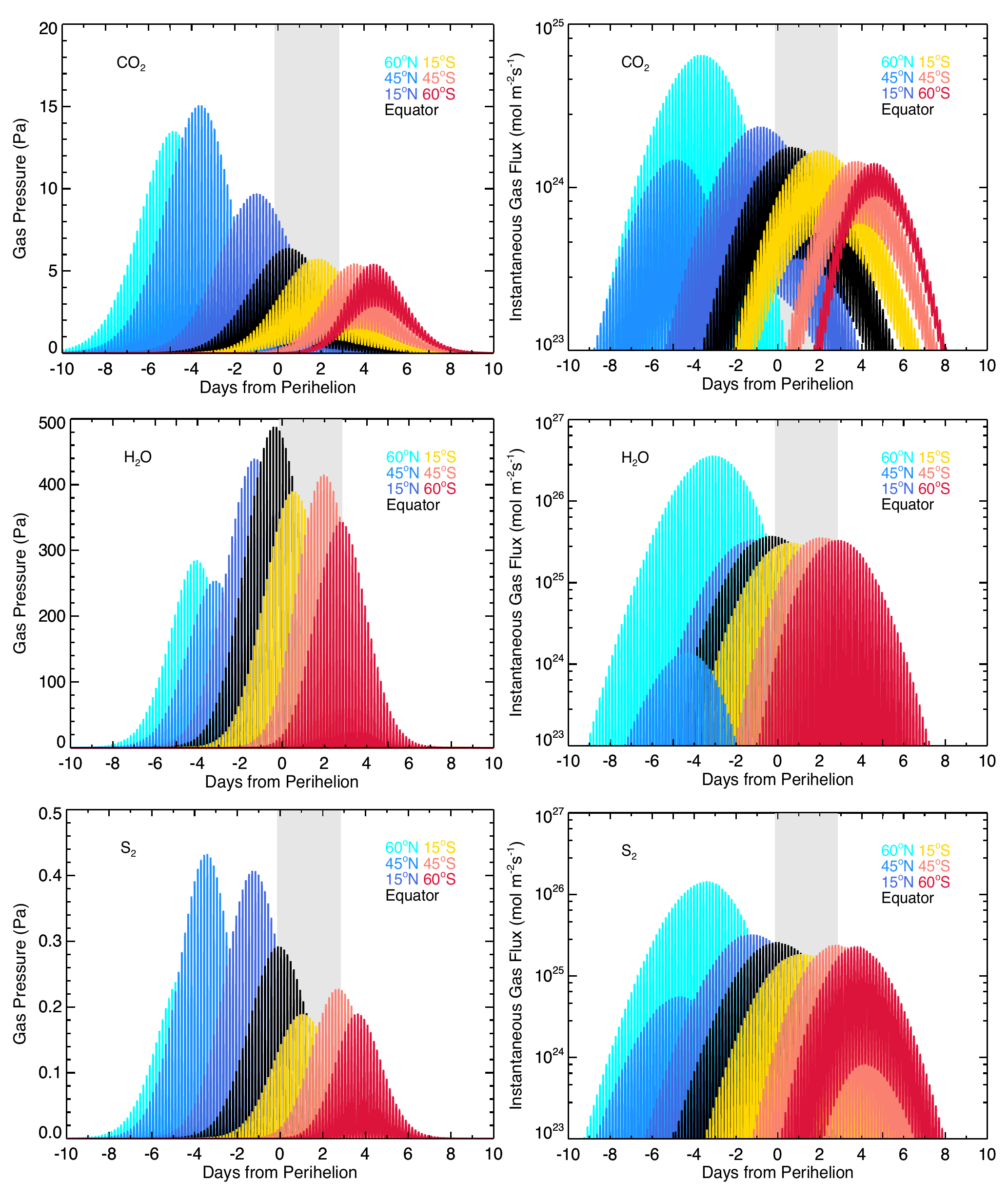}
	\caption{Left panels: Estimated subsurface gas pressure as a function of time for CO$_2$, H$_2$O, and S$_2$ produced at depths where only 1\% of the material decomposes each perihelion passage. Right panel: Instantaneous gas-flux estimates (molecules $\meter^{-2}$ $\second^{-1}$) for each gas species integrated through depths where up to 1\% of decomposition takes place.}\label{fig5}
\end{figure}

When computing the production rates and pressures it is difficult to accurately know the initial (pre-alteration) mineral abundances of the precursor material. Considering both our linear mixing model results and the measured composition of CY meteorites, we assume that 1\%, 20\%, and 30\% of the precursor material is comprised of carbonates, phyllosilicates, and sulfides, respectively \citep{King_etal19}. We select the depths where up to 1\% of each respective mineral experiences decomposition over a single perihelion passage (Extended \autoref{figex2}). Our estimated gas production rates shown in \autoref{fig5} are estimated for discrete latitudes, and \emph{total} gas production estimates could be up to $\sim10^8$ times greater, given that Phaethon's surface area is nearly 100 $\km^2$ \citep{MacLennan_etal22}. This activity is capable of ejecting dust from the surface of Phaethon, which can explain the dust trail associated with Phaethon's orbit \citep{Battams_etal22}.

\section*{Discussion}\label{sec2}

We identify the thermally-metamorphosed CY chondrites as the meteorite analog to Phaethon, ruling out previously suggested links to CK and CV chondrites \citep{Clark_etal10,deLeon_etal12,Masiero_etal21}. Other studies have proposed naturally and artificially heated meteorites as spectral analogs to Phaethon's abnormally blue-sloped reflectance spectrum \citep{Licandro_etal2007,Lazzarin_etal19,Kareta_etal21}. Notably, the CYs Y-86720 and Y-82162 have been identified as possible analogs \citep{Licandro_etal2007,Lazzarin_etal19}. These two meteorites exhibit clear similarities to Phaethon in the mid-infrared \autoref{fig2} as well as visible and near-infrared (Extended Data \autoref{figex3}). In particular, the abnormally blue slope of Phaethon's reflectance spectrum is consistent with CY meteorite chips.

Phaethon's mid-infrared spectrum does not show any strong transparency features that would be expected as a result of volumetric scattering from fine-grained silicates \citep{Salisbury_etal92,Hamilton10,Lane_etal11,Bates_etal21}, supporting the conclusion that Phaethon's northern hemisphere is boulder-rich or dominated by coarse regolith \citep{MacLennan_etal22}.
The $\sim10\%$ absolute reflectivities of CY chips at a wavelength of $0.55\um$ are systematically brighter than their powdered counterparts (see \autoref{supp}), and are consistent with Phaethon's geometric albedo of 11\% \citep{Geem_etal22}. This albedo value is within lower end of the albedo distributions of Pallas collisional family members ($0.10 < p_V < 0.18$) and is somewhat higher than other B-types ($0.03 < p_V < 0.10$) \citep{AliLagoa_etal16}. We posit that Phaethon's albedo is elevated from typical B-type values---rather than lower compared to Pallas family members---due to both mineralogical changes from thermal alteration \citep{Lindgren_etal20,Matsuoka_etal22} and having large grains on the surface \citep{MacLennan_etal22}. A connection to B-types outside of the Pallas family is consistent with the idea that Phaethon originated from an inner-belt B-type family \citep{MacLennan_etal21}. Mid-infrared spectra of primitive asteroid families in the inner asteroid belt will provide insight into Phaethon's origin and relationship to other B-types.

The best-fit olivine component in the spectral mixing model is Fo$_{75}$. We found that substituting spectra of other olivines in the mixing model yielded acceptable but less likely fits when using different olivine compositions (Extended Data \autoref{extab1}.). However, we also note that the position of emissivity features are consistent with band positions of natural and synthetic olivines with compositions in the range Fo$_{65-80}$ \citep{Hamilton10,Lane_etal11}. This range in olivine compositions is highly consistent with the bulk olivine in the CY meteorites Y-82162 (Fo$_{84-76}$) and Y-86029 (Fo$_{84-69}$) \citep{Tonui_etal03}, and the secondary olivine identified in Dhofar 1988 \citep[Fo$_{88-68}$;][]{Suttle_etal21}.

We find that the abundance of olivine ($35.6\%$) in the spectral mixing model is roughly consistent with its high abundances in CY chondrites \citep{King_etal19,Suttle_etal21}. However, the estimated abundance of troilite in the spectral mixing model ($41.8\%$) is roughly double than that of the average Fe-sulfide abundance among CY chondrites ($\sim 21\%$) \citep{King_etal19,Schrader_etal21}; this mismatch may be a limitation of the mixing model assumptions,  an indication of compositional heterogeneity on Phaethon, or both. The relative abundance of calcite to magnesite is consistent with meteoritic dolomite \citep{Tonui_etal03,Tonui_etal14}. Given that carbonates are uncommon in CY meteorites, the estimated abundance in the best-fit spectral model is relatively large ($\sim 7$ vol.\%). We posit that a significant amount of carbonates exist as veins on Phaethon's surface \citep[e.g., like those seen on Bennu;][]{Kaplan_etal20}, and are mostly destroyed in the decomposition process. This scenario can explain why they are not preserved in meteorite samples; alternatively, it is possible that the CY meteorites are not derived from carbonate-rich areas on Phaethon.

Hydroxides such as brucite and portlandite could have formed through the rehydration of the oxide products (i.e., MgO and CaO) of carbonate decomposition \citep[e.g.,][]{Haberle+Garvie17}. This scenario is possible if water vapor is sourced from partly dehydrated phyllosilicates in Phaethon's subsurface \citep{Nakato_etal08} during perihelion heating (Figure 5). Because the sheet-like structure of brucite is similar to phyllosilicates, it is possible that the emissivity feature at $\sim7\um$ is indicative of a partly decomposed phyllosilicate structure \citep{MacKinnon+Buseck79}. Portlandite that is found in the thermally-altered Sutter's Mill carbonaceous chondrite has been proposed to have formed via retroactive hydration upon cooling in the presence of water vapor \citep{Haberle+Garvie17}. Although Phaethon's surface is dehydrated, we suggest that the interior of Phaethon consists of relatively unaltered, hydrated CY ``precursor'' material that has not been heated. The periodic rise in temperature of deeply-buried phyllosilicates every perihelion can produce the water vapor needed for the rehydration of near-surface material.

We do not find evidence of enstatite or magnetite, which may be due to limitations rising from the noise in Phaethon's spectrum, and/or these minerals are actually not present near the surface. For example, any magnetite near the surface can react with sulfur gas to form Fe-sulfide \citep{Haberle+Garvie17}. Magnetite is proposed as an explanation for the blue spectral slopes of B-type asteroids \citep{Yang+Jewitt10} and is found in some CY meteorites \citep{King_etal19}, so we expect that it is present in low abundance on Phaethon and evades mid-infrared detection. Enstatite is lost during aqueous alteration, but can crystallize from decomposed phyllosilicates at temperatures $>1050\K$ \citep{Nakato_etal08,King_etal19,Suttle_etal21,Matsuoka_etal22}. The non-detection of enstatite pyroxene is consistent with their low abundance in CYs, yet may be present in low abundance on Phaethon \citep{Lisse+Steckloff22}.

The ejection of micron sized dust from the cracking of rock via thermal fracturing has been suggested as an explanation of Phaethon's activity \citep{Jewitt&Li2010}. However, the gas production from decomposition is perhaps a more efficient way of lifting dust from the surface given our estimates of gas pressure. It has recently been suggested that Phaethon's perihelion brightening and tail is not due to light scattering of dust, but instead a result of sodium photoemission near $589\nm$ \citep{Zhang23}. In the case that Phaethon's \emph{current} activity is indeed not caused by dust, we alternatively suggest that forbidden emission at $630\nm$ and $636\nm$ of oxygen derived from the photodissociation of H$_2$O or CO$_2$ gas as alternative explanation for the observed brightening. In any case, we claim that dust ejection via thermal decomposition gas production explains the presence of a dust stream associated with Phaethon's orbit \citep{Battams_etal22}. 

The extreme temperatures that Phaethon experiences during its perihelion passage will lead to the formation of secondary olivine via phyllosilicate decomposition (loss of all H$_2$O); e.g. for Lizardite:
\begin{equation}\label{eq:phyllo}
    \ce{ $\underset{\text{lizardite}}{\ce{(Mg,Fe)3Si2O5(OH)4}}$  ->  $\underset{\text{olivine}}{\ce{(Mg,Fe)2SiO4}}$ + $\underset{\text{silica}}{\ce{SiO$_2$}}$ + $\underset{\text{oxide}}{\ce{(Mg,Fe)O}}$ + 2H2O_{(g)}}
\end{equation}
Sulfide decomposition occurs above $\sim$970~K \citep{Hu_etal02} via the proposed reaction that releases sulfur gas:
\begin{equation}\label{eq:sulf}
    \ce{$\underset{\text{pyrrhotite}}{\ce{2FeS_{x}}}$ -> $\underset{\text{troilite}}{\ce{2FeS}}$ + (1-x)S_2_{(g)}},
\end{equation}
where $1 < x < 1.2$. Sulfur gas is also released when troilite decomposes into pure Fe metal at higher temperatures, but we only consider pyrrhotite decomposition for simplicity. When excess H$_2$O or O$_2$ is present, other species such as H$_2$S and SO$_2$ can theoretically form in the subsurface, which could then react with residual oxides to produce additional sulfides such as oldhamite (CaS) \citep{Haberle+Garvie17}. This scenario may be likely, as dehydration occurring deep in Phaethon's interior would means that some water vapor may be present in the upper regolith during perihelion passage.

Thermal decomposition of carbonates in meteorites has been observed at temperatures in the range $870\K-1070\K$ \citep{King_etal15}, and occurs via the reaction \citep[and references therein]{Kohobhange_etal19}:
\begin{equation}\label{eq:carb}
    \ce{$\underset{\text{carbonate}}{\ce{(Ca,Mg,Fe)CO3}}$ -> $\underset{\text{oxide}}{\ce{(Ca,Mg,Fe)O}}$ + CO_2_{(g)}}.
\end{equation}
This process, which releases CO$_2$ may explain the presence of Fe,Mg periclase (oxide) clasts in Y-86029 \citep{Tonui_etal03}. In the right conditions, the rehydration of magnesium oxide can result in brucite formation. We suppose that hydrated phyllosilicates can survive deep below Phaethon's surface for extended periods of time \citep{MacLennan_etal21}. The periodic heating of these phyllosilicates every perihelion can produce the water vapor needed for the rehydration of decomposition products to form hydroxides such as brucite and portlandite. For example, the rehydration of an oxide from \autoref{eq:carb} can occur via:
\begin{equation}
    \ce{ $\underset{\text{carbonate}}{\ce{(Ca,Mg)CO3_{(s)}}}$ + $\underset{\text{oxide}}{\ce{(Ca,Mg)O}}$ + H2O{(g)} ->
    $\underset{\text{hydroxide}}{\ce{2(Ca,Mg)(OH)2}}$ + CO2_{(g)} + O2_{(g)}}.
\end{equation}
Given that the decomposition reactions discussed above occur at somewhat different temperatures, Phaethon's subsurface could consist of layered strata representing material that has undergone various heating stages \citep{Nakamura05}. Gas produced from the various decomposition reactions can interact with solid material and form exotic minerals, such as those identified in heated primitive meteorites \citep{Haberle+Garvie17,Lindgren_etal20}.

Our estimated instantaneous production rates of CO$_2$, H$_2$O, and S$_2$ gas ($\sim10^{24}-10^{26}\ \textrm{mol} \second^{-1}$; Figure 5) exceed the estimated \emph{total} Fe production rate (10$^{22}$ atoms $\sec^{-1}$) \citep{Lisse+Steckloff22} and are somewhat smaller than the Na production estimate ($10^{26}$ atoms $\sec^{-1}$) \citep{Masiero_etal21}. Taking into account the surface area of Phaethon, the estimated gas production rates can reach $10^{32}-10^{34}\ \textrm{mol} \second^{-1}$, which is orders of magnitude larger than the Fe and Na estimates. Although our model is relatively simple when compared to these previous works, our compositional constraints that are based on CY meteorite analogs and the spectral mixing model provide a more realistic representation of the underlying activity mechanism. In any case, a more sophisticated model that accounts for, e.g., the effects of latent heat and vapor pressure \citep{Lisse+Steckloff22,Masiero_etal21} on the decomposition rates is warranted. Furthermore, a more general model that incorporates thermal decomposition and alteration history will be useful in assessing the long term (several $\kyr$) alteration and survivability of near-Sun asteroids \citep{2016Natur.530..303G,MacLennan_etal21}.

The DESTINY$^+$ mission will measure the dust environment and observe Phaethon's surface properties during a close flyby encounter \citep{Arai_etal2018}. Bright carbonate veins, similar to those identified on the surface of Bennu \citep{Kaplan_etal20}, and markers of thermal alteration, if present, may be detectable in the DESTINY$^+$ images of Phaethon's surface. Thermophysical modeling results which are consistent with polarimetric observations suggest surface heterogeneity on Phaethon \citep{MacLennan_etal22}. The heterogeneity could be leftover remnant of aqueous alteration processes, or a result of extreme heating and thermal decomposition. For example, the sudden temperature changes among the southern latitudes during perihelion passage can lead to more efficient breakdown of the surface, compared to the gradual heating of northern latitudes before perihelion. By this same logic, we also expect latitudinal banding of material that has been heated to different maximum temperatures on Phaethon.

The extended mission target of DESINY$^+$ is (155140) 2005 UD, which is thought to share a common origin with Phaethon due to its similar orbital and physical properties \citep[and references therein]{Devogele_etal20}. However, doubts about this compositional link were raised by newly-observed near-infrared spectral slope differences between the pair \citep{Kareta_etal21}. We point out that spectral slope differences correlated with grain sizes can be seen among CY powders and chips (Extended Data \autoref{figex3}). Mismatches in the spectral slope can be explained by the particle size variations and are not necessarily indicative of compositional differences. Given the similar spectroscopic and polarimetric properties to Phaethon, we therefore predict that 2005~UD is also comprised of CY material and can be verified via observing its mid-IR spectrum.

\section*{Data \& Methods}\label{sec3}

\subsection*{Emissivity Spectrum}
The mid-IR spectrum of Phaethon were acquired by the Spitzer Space Telescope Infrared Spectrograph \citep{Houck_etal04,Hanus_etal2016}. In order to calculate an emissivity spectrum, we divide the observed fluxes by a modeled continuum. Continuum fluxes are generated from the TPM by fixing the rotation phase of a shape model for the midpoint time of the observation and using the previously obtained best-fit model parameters from MacLennan et al. (2022) \citep{MacLennan_etal22} ($D_\mathit{eff} = 5.4\km$, $\Gamma = 480 \tiu$, and low default roughness) for the dataset. We employ the \texttt{loess} (locally estimated scatterplot smoothing) algorithm in \texttt{R} to generate a smoothed function for use in the PCA (next subsection); a 2$nd$ degree polynomial with a degree of smoothing of 0.7 is used.

\subsection*{Principal-Component Analysis}

We acquired several mid-infrared relfectance spectra of meteorites from the RELAB database. The wavelength ($\lambda$) dependent spectral reflectance values ($R$) are converted to emissivity ($\varepsilon$) via Kirchhoff's law of thermal radiation: $\varepsilon (\lambda) = 1-R (\lambda)$. Supplementary \autoref{tabsupp1} lists the RELAB file name for spectra used. A spectrum of Sutter's Mill (SM3) was also included \citep{Haberle_etal19}. We prepare each spectrum for the PCA by removing the overall spectral slope and re-sampling to the same wavelengths as the Phaethon IRS spectrum. Finally, we adjust the spectral contrast of each spectrum to unity in order to mitigate grain-size and porosity effects of the samples \citep[][and references therein]{Salisbury_etal92,Shirley+Glotch19}.

\subsection*{Linear Spectral-Mixing Model}

Mineral spectra (Supplementary \autoref{tabsupp2}) are fit to Phaethon's spectrum using the linear model function \texttt{lm} in the \texttt{R}. We also simulate the presence of a neutral component in order to correct for differences in spectral slope among these spectra. Data in the wavelength range $5.55\um-35\um$ are used, which is a limitation based on the wavelength coverage of the library spectra, but also excludes portions of Phaethon's spectrum with the most noise. The mineral spectra are left unaltered, except for cases in which a spectrum does not span the entire wavelength range. Additionally, we found that the Fo75 spectrum in the Lane et al. (2011) \citep{Lane_etal11} suite exhibited a larger spectral range compared to the other compositions. We scaled this spectrum in order to reduce the spectral contrast by 30\% which was more comparable to the other olivine spectra.  We found that this affected the relative abundance of troilite and olivine in the mixing model: without this correction, the abundance of troilite and olivine is $\sim$10\% larger and lower, respectively, compared to our result (Table 1). With this demonstration, and because we don't know the particle sizes of each individual mineral component, we suppose that the reported uncertainties in their modeled abundances as described below could be underestimated by a factor of a few.

We begin with all spectral members included in the model, and remove components with negative coefficients, one at a time, until all remaining components possess positive coefficients. This approach is often referred to as non-negative least squares methods \citep{Ramsey+Christensen98}. We also tested a stepwise linear model in \texttt{R} that uses the Bayesian Information Criterion, $BIC$, to assess and remove non-significant coefficients (components). Both approaches yielded the same unique result. The best-fit component mixture yields a residual root-mean squared (r.m.s.) error of 0.0114, and a reduced chi-squared goodness of fit of $\chi^2_\nu = 0.837$. We also computed the difference in the Bayesian Information Criterion, $\Delta BIC$, between all models and the best-fit and found that all are larger than the $\Delta BIC > 6$ used to favor one model over another \citep{Kass&Raferty95}. Model coefficients and normalized abundances are given in Table 1, along with variations of the nominal model with individual species removed.

We investigated the effect of different carbonate species on the model fit by substituting in siderite (FeCO$_3$) and ankerite (Ca(Fe,Mg,Mn)(CO$_3$)$_2$). Replacing magnesite with siderite or ankerite did not change the r.m.s. value significantly and $\chi^2_\nu<1$ were acceptable. These alternative combinations required similar carbonate abundances as the nominal model ($\sim 5-8 vol. \%$). Therefore, we suppose that any of these carbonates could exist on Phaethon. We also considered spectra of pyrrhotite and pyrite (FeS$_2$) from the CRISM spectral library: The spectrum of pyrite is completely featureless and the inclusion of pyrrhotite did not significantly improve the fit. Therefore, we suppose that troilite is the dominant sulfide component in Phaethon's spectrum.

\subsection*{Thermal Decomposition Model}\label{sub:temp}

The \texttt{orbTPM} thermophysical model of MacLennan, et al. (2021) \citep{MacLennan_etal21} is used to calculate Phaethon's surface temperatures. This model solves the one dimensional heat diffusion equation and computes surface and sub-surface temperatures of a spherical object. The rotation period of Phaethon is used ($3.603955 \hour$) with the spin axis of the nonconvex shape model (ecliptic longitude and latitude of 316$\deg$ and -48.7$\deg$, respectively) along with a thermal inertia of $\Gamma = 480 \tiu$ \citep{MacLennan_etal22}. Consequently, the thermal skin depth is $l_s = 34.5^{+6.3}_{-4.1} \mm$, when using a heat capacity of $c_s = 773 \J \kg^{-1} \K^{-1}$, a solid density of $2800 \kg \meter^{-3}$ \citep{Macke_etal11} and porosity of $\phi = 0.4$. Temperatures are calculated for 13 discrete latitudes separated by 15$\deg$ ranging from $-90\deg$ to $+90\deg$. Phaethon's fast rotation rate poses some difficulty when surface temperatures become very large near perihelion; thus, we use 1800 time steps per one rotation in order to ensure model convergence and stability.


In a first-order decomposition reaction, $A_{(s)} \rightarrow B_{(s)} + C_{(g)}$, the instantaneous change in the amount of reactant ($dA$, where $A$ is e.g., carbonate, sulfide, or phyllosilicate and $C$ is the gas species) is dependent on the amount of reactant that has not decomposed at a given time $A(t)$ and the rate constant:
\begin{equation}\label{eq:reac}
     \frac{dA}{dt} = -A(t) k_0.
\end{equation}
The rate constant, $k_0$ ($\second^{-1}$), of the first-order thermal decomposition reaction as a function of temperature, $T$, can be calculated via the Arrhenius equation:
\begin{equation}\label{eq:kconts}
    k_0 = A_0\exp{\Big(\frac{-E_a}{R_0 T}\Big)}
\end{equation}
where $R_0 = 8.314 \J \K^{-1} \mol^{-1}$, with the activation energy, $E_a$, and the pre-exponential factor, $A_0$ ($\second^{-1}$) experimentally determined (Supplementary \autoref{tabsupp3}). The $E_a$ and $A$ values vary based on the experimental setup and variations in the sample, and the values we use are approximate averages. We find that the differences in $E_a$ between different carbonates are on the same order of variation in $E_a$ due to particle size effects (i.e., $170-210$kJ mol$^{-1}$ for calcite \citep{RodriguezNavarro09}). Because we use a typical value in this work, we suppose that our value of $E_a$ is suitable for an order of magnitude estimation of the gas flux.

We use a time step of $\Delta t = 61.8 \second$ and \autoref{eq:reac} to estimate the decrease in reactant as a function of time. The amount of gas produced at each time step is determined from the change in reactant:
\begin{equation}
    \Delta \rho_{gas} = -f_{conv}\ \Big(\frac{1}{\phi}-1\Big)\ \Delta \rho_{solid}
\end{equation}
in which $f_{conv}$ is the mass fraction of reactant that is converted into gas and $\rho_{solid}$ is calculated from \autoref{eq:reac}. Approximately $f_{conv} = 43\%$ of carbonate is converted into CO$_2$ gas (\autoref{eq:carb}), $f_{conv} =11.9\%$ of phyllosilicate into H$_2$O gas (\autoref{eq:phyllo}), and $f_{conv} = 23\%$ of sulfide into sulfur gas (\autoref{eq:sulf}). We assume a density of the solid material that of $2800 \kg \meter^{-3}$ \citep{Macke_etal11}, meaning that the initial densities are: $\rho_{carb} = 28 \kg \meter^{-3}$, $\rho_{phyllo} = 560 \kg \meter^{-3}$, and $\rho_{carb} = 840 \kg \meter^{-3}$. The volume ratio of void space to solid space in a porous medium is $(1/\phi - 1)$ where we assume a porosity of $\phi = 0.4$.

The gas pressure is calculated from the ideal gas law and gas density from:
\begin{equation}
    P_{gas} = \frac{\rho_{gas}}{m_{mol}}R_0 T.
\end{equation}
Finally, the mass flux ($\Phi$) at a particular depth is calculated from $P_{gas}$ and the mean thermal velocity ($v_{th}$):
\begin{equation}
    \Phi = P_{gas} v_{th} = P_{gas} \sqrt{\frac{8R_0T}{\pi m_{mol}}}.
\end{equation}

\backmatter

Correspondence and material requests should be addressed to Eric MacLennan (eric.maclennan@helsinki.fi).

\section*{Acknowledgments}

This work is based in part on observations made with the Spitzer Space Telescope, operated by the Jet Propulsion Laboratory, California Institute of Technology under a contract with NASA. The authors thank Joshua P. Emery for sharing the previously published Spitzer-IRS spectrum of Phaethon.







\section*{Availability of data and materials:}
All datasets analyzed in this study are publicly available. The mid-infrared flux spectrum of Phaethon measured by the Spitzer Space Telescope's Infrared Spectrograph instrument (AOR: 4890624) is publicly available on the Spitzer Heritage Archive (\url{https://sha.ipac.caltech.edu/}). The reduced spectrum was published in Hanu\v{s}, et al. (2016) \citep{Hanus_etal2016} and shared with the author(s) from Joshua P. Emery, a co-author of that study. Laboratory spectra of meteorites and minerals (Supplementary \autoref{tabsupp2}) can be downloaded from the RELAB databse via the PDS Geosciences Node Spectral Library (\url{https://pds-speclib.rsl.wustl.edu/}), ASU (\url{https://speclib.asu.edu}), and CRISM spectral libraries (\url{http://speclib.rsl.wustl.edu/}).

\section*{Code availability:}
The \texttt{orbTPM} code is available in GitHub in \url{https://github.com/cosmicdustbeing/asteroid-thermal}. R packages (stats) are publicly available for download.

\section*{Authors' contributions:}
E. M. conceived the original idea, carried out the analysis, interpreted the results, and was lead writer in the manuscript. M. G. assisted with the interpretation of results and provided consultation for the manuscript.
Both authors discussed the results and contributed to the final manuscript.

\section*{Tables}

\begin{table}[h!]
    \centering
    \caption{Best-fit Linear Spectral Mixing Model Components}
    \begin{tabular}{l|ccccr}
        mineral & coefficient & norm. abundance & $^a\Delta \chi^2_\nu$ \\ \hline
        olivine (Fo$_{75}$) & 0.301 $\pm$ 0.025 & 35.6 $\pm$ 2.5\% & 0.663 \\
        troilite & 0.353 $\pm$ 0.027 & 41.8 $\pm$ 3.5\% & 0.794 \\
        portlandite & 0.055 $\pm$ 0.018 & 6.6 $\pm$ 2.1\% & 0.028 \\
        calcite & 0.036 $\pm$ 0.004 & 4.2 $\pm$ 0.5\% & 0.210 \\
        magnesite & 0.027 $\pm$ 0.005 & 3.2 $\pm$ 0.6\% & 0.067 \\
        brucite & 0.073 $\pm$ 0.012 & 8.7 $\pm$ 1.4\% & 0.099 \\ \hline
        $\log_{10}(\lambda)$ & 0.054 $\pm$ 0.003 & --- & --- \\ \hline
        \multicolumn{4}{l}{$^a$ increase in $\chi^2_\nu$ when removed from the mixing model.}
    \end{tabular}
\end{table}

\section*{Figures}

\clearpage

\section*{Extended Data}

\renewcommand{\figurename}{Extended Figure}
{
\setcounter{figure}{0}
\begin{figure}[h!]
	\centering
	\includegraphics[width=0.9\linewidth]{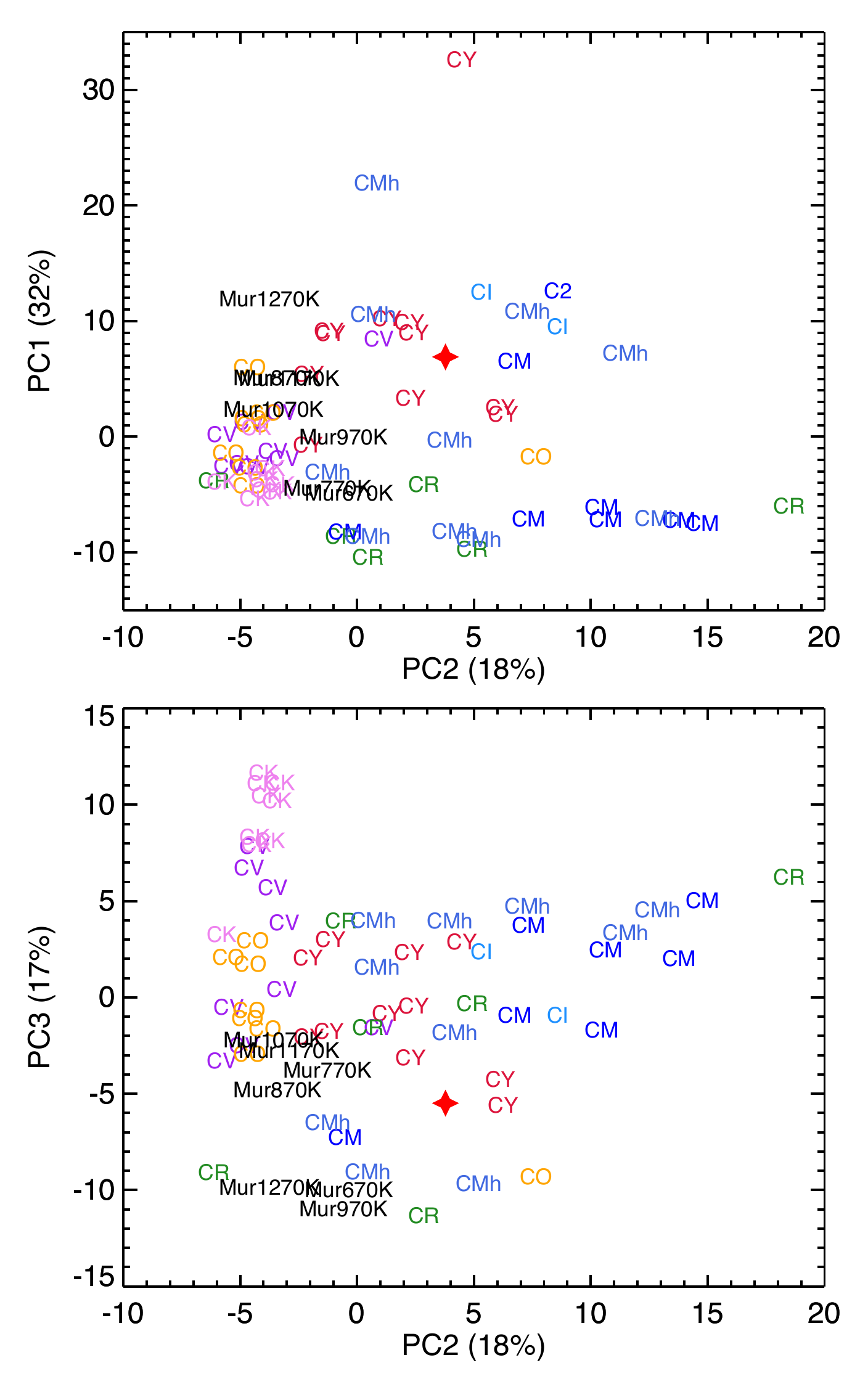}
	\caption{Principal components (PC\#) of mid-infrared spectra of carbonaceous chondrites. Datapoints are labeled and colored by their meteorite group with ``CMh'' indicating naturally heated carbonaceous chondrites and ``Mur'' corresponding to samples of the Murchison meteorite heated to the indicated temperature. Phaethon is indicated by the red star.}\label{figex1}
\end{figure}
}

\renewcommand{\figurename}{Extended Figure}
{
\begin{figure}[h!]
	\centering
	\includegraphics[width=0.7\linewidth]{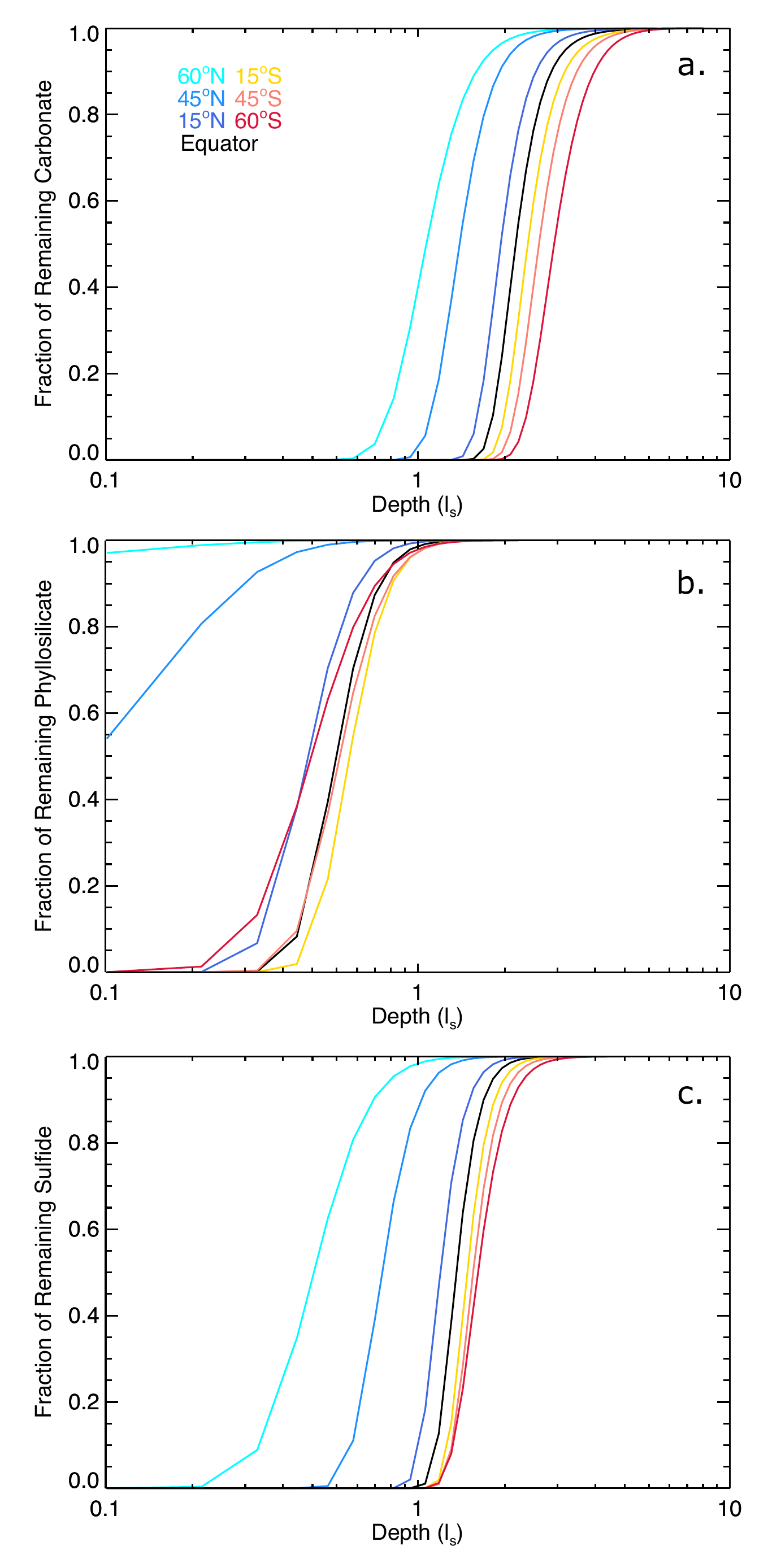}
	\caption{Fraction of decomposed material (carbonate, phyllosilicate, or sulfide in panels a., b., and c., respectively) after one perihelion passage as a function of depth beneath the surface, expressed in terms of the thermal skin depth ($l_s$) \citep{MacLennan_etal22}.}\label{figex2}
\end{figure}
}

\renewcommand{\figurename}{Extended Figure}
{
\begin{figure}[h!]
	\includegraphics[width=\linewidth]{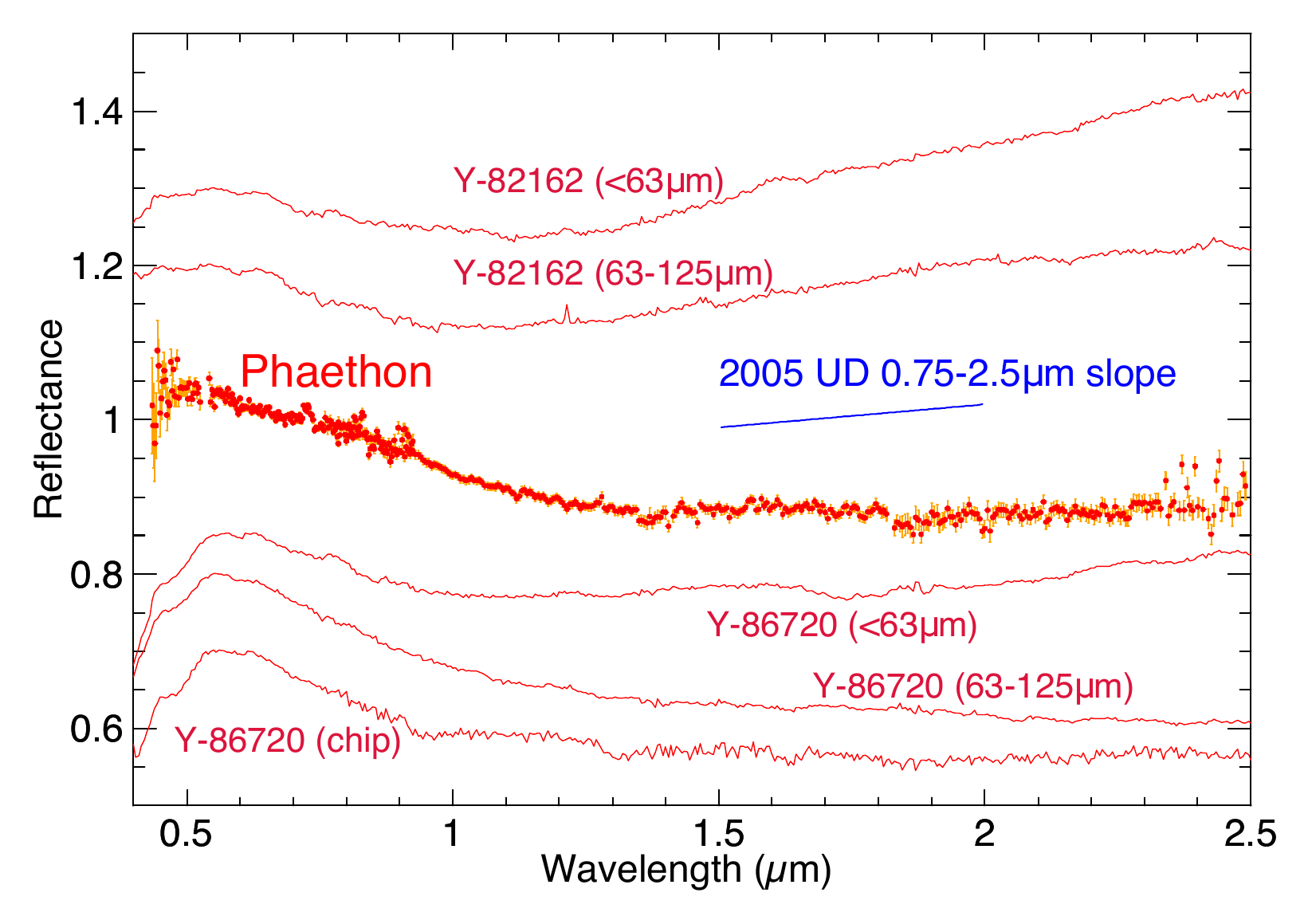}
	\caption{Visible and near-infrared reflectance spectra of Phaethon and two CY meteorites ground into different size fractions or left  as whole rock ''chips''. The 1$\sigma$ uncertainty for each reflectance value are shown as orange vertical bars. The blue line represents the average near-infrared ($0.75 - 2.5 \um$) spectral slope of 2005 UD, as observed by Kareta et al. (2021) \citep{Kareta_etal21}.}\label{figex3}
\end{figure}
}

\setcounter{table}{0}
\renewcommand{\tablename}{Extended Data Table}
{
\begin{table}[h!]
\caption{Spectral Mixing Models with Different Olivine and Carbonate Components}\label{extab1}
    \begin{tabular}{l|ccc}
        mineral mixture & r.m.s. & $\chi_\nu^2$ & $\Delta BIC$ \\ \hline
        Fo$_{75}$ + troi + port + cal + magn + bru & 0.0115 & 0.840 &  -- \\
        Fo$_{75}$ + troi + cal + magn + bru (- port) & 0.0119 & 0.897 &  15 \\
        Fo$_{75}$ + troi + port + cal + bru (- magn) & 0.0119 & 0.905 &  18 \\
        Fo$_{75}$ + troi + port + cal + magn (- bru) & 0.0126 & 0.995 &  51 \\
        Fo$_{75}$ + troi + port + magn + bru (- cal) & 0.0128 & 1.023 &  61 \\
        troi + port + cal + magn + bru (- Fo$_{75}$) & 0.0141 & 1.464 &  142 \\
        Fo$_{75}$ + port + cal + magn + bru (- pyrr) & 0.0157 & 1.630 &  195 \\
        \hline
        Fo$_0$ + troi + port + cal + magn + enst + bru & 0.0136 & 1.223 & 120 \\
        Fo$_{10}$ + troi + port + sid + magn + enst + peri & 0.0131 & 1.067 & 90 \\
        Fo$_{20}$ + troi + port + sid + magn + enst + peri & 0.0127 & 1.005 & 68 \\
        Fo$_{30}$ + troi + port + sid + magn + enst + peri & 0.0123 & 0.961 & 51 \\
        Fo$_{40}$ + troi + port + sid + magn + enst + bru & 0.0122 & 0.962 & 47 \\
        Fo$_{50}$ + troi + port + sid + magn + bruc & 0.0118 & 0.884 & 19 \\
        Fo$_{55}$ + troi + port + cal + magn + enst + bru & 0.0119 & 0.893 & 27 \\
        Fo$_{65}$ + troi + port + cal + magn + enst + bru & 0.0116 & 0.838 & 11 \\
        Fo$_{70}$ + troi + port + cal + magn + bru & 0.0120 & 0.944 & 36 \\
        Fo$_{80}$ + troi + port + cal + magn + enst + bru & 0.0118 & 0.911 & 30 \\
        Fo$_{89}$ + troi + port + cal + magn + enst + bru & 0.0128 & 1.089 & 83  \\
        Fo$_{100}$ + troi + port + cal + magn + enst + bru & 0.0136 & 1.336 & 131 \\
        \hline       
        Fo$_{75}$ + troi + port + cal + ank + bru & 0.0117 & 0.884 & 17 \\
        Fo$_{75}$ + troi + port + cal + sid + bru & 0.0117 & 0.869 & 14 \\
        Fo$_{75}$ + troi + port + magn + ank+ bru & 0.0118 & 0.890 & 21 \\
        Fo$_{75}$ + troi + port + magn + sid + bru & 0.0117 & 0.869 & 16 \\
        Fo$_{75}$ + troi + port + ank + sid + bru & 0.0118 & 0.876 & 19 \\

        \hline
        \multicolumn{4}{l}{pyrr - pyrrhotite, Fo - forsterite, bru - brucite, enst - enstatite, port - portlandite,} \\
        \multicolumn{4}{l}{cal - calcite, magn - magnesite, sid - siderite, ank - ankerite, peri - periclase}
    \end{tabular}
\end{table}
}

\clearpage

\section*{Supplementary Information }\label{supp}

\setcounter{table}{0}
\renewcommand{\tablename}{Supplementary Data Table}
{
\begin{table}[h!]
\caption{List of RELAB sample spectra used}\label{tabsupp1}
    \begin{footnotesize}
    \begin{tabular}{lcl||lcl}
        Name, Sample & Type & RELAB ID & Name,Sample & Type & RELAB ID \\ \hline
        A-880835,59 & CV3 & MP-TXH-177 & GRA 06100,94 & CR2 & MP-TXH-254 \\
        A-881317,75 & CV3 & MP-TXH-178 & GRO 03116,48 & CR2 & MP-TXH-255 \\
        ALH 84028,85 & CV3 & MC-RPB-005 & GRO 95577,84 & CR1 & MP-TXH-256 \\
        ALH 85006,51 & CV3 & MP-TXH-261 & MIL090001,73 & CR2 & MP-TXH-231 \\
        Allende & CV3 & MT-BEC-092-B & Y-793495,87 & CR2 & MP-TXH-176 \\
        Allende & CV3 & MB-TXH-063-C & Y-790112,87 & CR2 & MP-TXH-175 \\
        Allende & CV3 & MT-BEC-092-A & Y-980115,101 & CY & MP-TXH-139 \\
        LAR 12002,23 & CV3 & MP-TXH-264 & Y-86720,100 & CY & MP-TXH-159 \\
        MIL 13328,9 & CV3 & MP-TXH-266 & Y-82162,98 & CY & MP-TXH-137 \\
        ALH 85002,111 & CK4 & MP-TXH-291 & Y-82162,79 & CY & MB-CMP-019-B \\
        A-881551,81 & CK6 & MP-TXH-161 & Y-82162,79 & CY & MB-CMP-019-C \\
        EET 87507,27 & CK5 & MB-TXH-092 & B-7904,116 & CY & MP-TXH-146 \\
        EET 87860,29 & CK5-6 & MP-TXH-303 & B-7904,116 & CY & MP-TXH-146 \\
        EET 92002,10 & CK5 & MC-RPB-003 & B-7904 & CY & MB-CMP-018-B \\
        EET 92002.65 & CK5 & MP-TXH-302 & B-7904 & CY & MB-CMP-018-C \\
        LEW 87009,64 & CK6 & MP-TXH-304 & EET 83226,22 & C2-un* & MP-TXH-211 \\
        Y-82102,95 & CK5 & MP-TXH-166 & EET83355,32 & C2-un* & MP-TXH-212 \\
        Y-82191,88 & CK6 & MP-TXH-165 & Jbilet Winselwan & CM2* & MT-DAK-312 \\
        ALH A77003,106 & CO3.6 & MP-TXH-171 & Murchison & CM2** & MB-TXH-064 \\
        ALH 82101,29 & CO3.4 & MP-TXH-268 & PCA 91008,31 & CM2* & MT-JMS-198 \\
        ALH 83108,38 & CO3.5 & MC-RPB-004 & QUE 93005 & CM* & MT-JMS-220 \\
        ALH 85003,41 & CO3.5 & MP-TXH-270 & WIS 91600,700 & CM-an* & MP-TXH-234 \\
        MET 00694,7 & CO3.6 & MP-TXH-273 & Y-82098,105 & CM2* & MP-TXH-156 \\
        MET 00737,15 & CO3.6 & MP-TXH-274 & Y-791191,95 & CM2* & MP-TXH-148 \\
        Y-81020,114 & CO3.0 & MP-TXH-174 & \\
        Y-791717,113 & CO3.3 & MP-TXH-173 & \\
        ALH 83100,196 & CM1/2 & MC-RPB-001 \\
        Alias & CI1 & MT-KTH-264 \\
        LEW 90500,45 & CM2 & MC-RPB-002 \\
        Murchison & CM2 & MB-TXH-064-D4 \\
        Murchison & CM2 & MT-S1S-234 \\
        Orgueil & CI1 & MT-JMS-191 \\
        Tagish Lake & C2-un & MT-MEZ-318-A \\
        Y-74662,109 & CM2 & MP-TXH-147 \\
        Y-86695,102 & CM2 & MP-TXH-158 \\ \hline
        \multicolumn{6}{l}{* Naturally Heated; ** Artificially Heated; ALH - Allan Hills; LAR - Larkman Nunatak;} \\
        \multicolumn{6}{l}{MIL - Miller Range; A - Asuka; EET - Elephant Moraine; LEW - Lewis Cliff; Y - Yamato;} \\
        \multicolumn{6}{l}{MET - Meteorite Hills; GRA - Graves Nunataks; GRO - Grosvenor Mountains;} \\
        \multicolumn{6}{l}{PCA - Pecora Escarpment; QUE - Queen Alexandra Range}
    \end{tabular}
\end{footnotesize}
\end{table}
}

\renewcommand{\tablename}{Supplementary Table} 
{
\begin{table}[h!]
\caption{Sources of all mineral spectra considered in the spectral mixing model}\label{tabsupp2}
    \begin{tabular}{lc|c}\footnotesize
        mineral & formula & library ID/reference \\ \hline
        ankerite & Ca(Fe,Mg,Mn)(CO$_3$)$_2$ & ASU \# 2290 \\
        brucite & Mg(OH)$_2$ & Glotch \& Rogers (2013) \\
        calcite & CaCO$_3$ & Glotch \& Rogers (2013) \\
        enstatite & MgSiO$_3$ & ASU \#536 \\
        olivine (Fo$_{0,...,100}$) & (Fe,Mg)SiO$_2$ & ASU, Lane et al. (2011) \\
        lime & CaO & Glotch \& Rogers (2013) \\
        magnesite & MgCO$_3$ & Glotch \& Rogers (2013) \\ 
        magnetite & Fe$^{2+}$Fe$^{3+}_2$O$_4$ & Glotch et al. (2004) \\
        periclase & MgO & Glotch \& Rogers (2013) \\
        pigeonite & (Ca,Mg,Fe)(Mg,Fe)Si$_2$O$_6$ & ASU \#3668 \\
        portlandite & Ca(OH)$_2$ & Glotch \& Rogers (2013) \\
        pyrrhotite & FeS$_x$ ($1 < x < 1.2$) & Donaldson-Hanna et al. (2021) \\
        saponite & Ca$_{0.25}$(Mg,Fe)$_3$((Si,Al)$_4$O$_{10}$)(OH)$_2\cdot$n(H$_2$O) &ASU \#579 \\
        serpentine & (Si,Al,Fe)$_2$O$_5$(OH)$_4$ & ASU \#443 \\
        siderite & FeCO$_3$ & ASU \#530 \\
        troilite & FeS & Donaldson-Hanna et al. (2021) \\ \hline
        \multicolumn{3}{l}{{\bf Ref:} ASU (\url{https://speclib.asu.edu}); Donaldson-Hanna et al. (2021) \citep{DonaldsonHanna_etal21}; Glotch et al. (2004) \citep{Glotch_etal04};} \\
        \multicolumn{3}{l}{Glotch \& Rogers (2013) \citep{Glotch+Rogers13}; Lane et al. (2011) \citep{Lane_etal11}}
    \end{tabular}
\end{table}
}

\renewcommand{\tablename}{Supplementary Table}
{
\begin{table}[h!]
\caption{Reaction parameters for thermal decomposition model}\label{tabsupp3}
    \begin{tabular}{lc|ccr}
        mineral & gas species & E$_a$ (kJ$ \mol^{-1}$) & $A_0$ ($\second ^{-1}$) & ref. \\ \hline
        carbonate & CO$_2$ & 190 & $1.3 \times 10^9$ & \citep{RodriguezNavarro09} \\
        phyllosilicate & H$_2$O & 300 & $3.9 \times 10^{12}$ & \citep{Zhou_etal17} \\
        Fe-sulfide & S$_2$ & 270 & $1.8 \times 10^{13}$ & \citep{Hu_etal02}\\ \hline
    \end{tabular}
\end{table} 
}







\clearpage

\bibliography{bibfile} 


\end{document}